\documentclass[12pt]{iopart}

\usepackage{amsmath} \usepackage{amssymb}
\usepackage{color}
\usepackage{graphicx}

\newcommand*{\un}[1]{\ensuremath{\ \mathrm{#1}}}
\newcommand{\trace}{\ensuremath{\mathrm{tr}}} 

\newcommand{\ket}[1]{\ensuremath{\left | #1 \right \rangle}} 
\newcommand{\bra}[1]{\ensuremath{\left \langle #1 \right |}} 
\newcommand{\mean}[1]{\ensuremath{\left \langle #1 \right \rangle}} 
\newcommand{\diff}{\ensuremath{\mathrm{d}}} 
\newcommand{\spvec}[1]{\ensuremath{\mathbf{#1}}} 

\newcommand{\partdiff}[2]{\ensuremath{\frac{\partial #1}{\partial
#2}}} 

\newcommand{\FuncD}[1]{\ensuremath{\mathcal{D}#1}} 

\newcommand{\hFuncDeriv}[2]{\ensuremath{\frac{\hbar \delta #1}{\delta #2}}}

\newcommand{\qop}[1]{\ensuremath{\hat{#1}}}

\newcommand{\EMscalar}[0]{\ensuremath{\varphi}} 
\newcommand{\vs}[0]{\ensuremath{v_{\text{S}}}}
\newcommand{\vsvec}[0]{\ensuremath{\spvec{\vs}}} 

\newcommand{\vcrit}[0]{\ensuremath{v_{\text{crit}}}}
\newcommand{\dvsvec}[0]{\ensuremath{\delta \vsvec}} 

\newcommand{\sfdens}[0]{\ensuremath{\rho_{\text{S}}}} 

\newcommand{\Ek}[1]{\ensuremath{E_{#1}}} 
\newcommand{\EkDeltaSq}[1]{\ensuremath{\Omega_{#1}}} 

\newcommand{\GIphase}[0]{\ensuremath{\gamma}}
\newcommand{\GIDphase}[0]{\ensuremath{\Delta\GIphase}} 
\newcommand{\ephase}[0]{\ensuremath{\phi}}
 
\newcommand{\FTmeasure}{\ensuremath{(2\pi)^3}}
\newcommand{\taumeasure}{\hbar} 
 
\newcommand{\identity}[0]{\ensuremath{\nambumat{1}}}
\newcommand{\nambuvec}[1]{\ensuremath{\boldsymbol{#1}}}
\newcommand{\nambumat}[1]{\ensuremath{\boldsymbol{#1}}} 

\newcommand{\ftrafo}[1]{\ensuremath{\tilde{#1}}}

\newcommand{\FTnambumat}[1]{\ensuremath{\ftrafo{\nambumat{#1}}}}

\newcommand{\GenFunc}[2]{\ensuremath{\mathcal{#1} \left [ #2 \right
]}}

\newcommand{\extflux}[0]{\ensuremath{\Phi_{\text{ext}}}}

\newcommand{\leftcurrket}[0]{\ensuremath{\ket{\circlearrowleft}}}
\newcommand{\rightcurrket}[0]{\ensuremath{\ket{\circlearrowright}}}
\newcommand{\leftcurrbra}[0]{\ensuremath{\bra{\circlearrowleft}}}
\newcommand{\rightcurrbra}[0]{\ensuremath{\bra{\circlearrowright}}}

\newcommand{\fermi}[0]{\ensuremath{\mu_{\text{F}}}}
\newcommand{\fermiv}[0]{\ensuremath{v_{\text{F}}}}

\newcommand{\Dj}[0]{\ensuremath{\delta
\spvec{j}}} 

\newcommand{\Dmeanvs}[0]{\ensuremath{\delta\!\mean{\vsvec}}}

\newcommand{\Dmeanjpos}[1]{\ensuremath{\delta\!\mean{\spvec{j}(#1)}}}

\newcommand{\length}[0]{\ensuremath{L}}  

\newcommand{\modeN}[0]{\ensuremath{N}} 
\newcommand{\DmodeN}[0]{\ensuremath{\Delta \modeN}}
\newcommand{\DNtot}[0]{\ensuremath{\Delta N_{\text{tot}}}}

\newcommand{\modeNk}[1]{\ensuremath{\modeN_{#1}}}

\newcommand{\DmodeNk}[1]{\ensuremath{\DmodeN_{#1}}}

\newcommand{\NumModes}[0]{\ensuremath{\mathcal{N}}}

\newcommand{\Ip}[0]{\ensuremath{I_{\text{p}}}} 
\newcommand{\DIp}[0]{\ensuremath{\delta \Ip}}

\newcommand{\Voltage}{\ensuremath{V}}

\newcommand{\modedensity}[1]{\ensuremath{\rho_{#1}}}
\newcommand{\DNdensity}[0]{\ensuremath{\Delta n}}

\newcommand{\dcostheta}[0]{\ensuremath{\diff(\cos \theta)}}
\newcommand{\DNkern}[0]{\ensuremath{\mathcal{K}}}
\newcommand{\Fermiq}[1]{\ensuremath{#1_{\text{F}}}}
\newcommand{\rhoF}[0]{\ensuremath{\modedensity{\text{F}}}}

\begin{document}

\title{Electronic structure of superposition states in flux qubits.}


\author{Jan I. Korsbakken$^{1,2}$, Frank K. Wilhelm$^3$ and K. Birgitta Whaley$^{2}$}

\address{$^1$Department of Physics and Berkeley Quantum Information and Computation Center,
, University of California, Berkeley,
94720-7300, U.S.A.}
 \address{$^2$Department of Chemistry and Berkeley Quantum Information and Computation Center, University of California, Berkeley,
94720-1460, U.S.A.}
\address{$^3$Institute for Quantum Computing and Department of Physics and
Astronomy, University of Waterloo, 200 University Avenue West,
Waterloo, ON, Canada, N2L 3G1}
\ead{whaley@berkeley.edu}

\begin{abstract}
Flux qubits, small superconducting loops interrupted by Josephson junctions, are
successful realizations of quantum coherence for macroscopic variables. Superconductivity in these loops is carried by $\sim 10^6$ -- $10^{10}$ electrons, which has been interpreted as suggesting that coherent superpositions of such current states are macroscopic superpositions analogous to
the state of Schr\"odinger's cat.  We provide 
a full microscopic analysis of such qubits, from which the macroscopic quantum description can be derived. This reveals that the number of microscopic constituents participating in superposition states for experimentally accessible flux qubits is surprisingly but not trivially
small. The combination of this relatively small size with large differences between macroscopic observables in the two branches is seen to result from the Fermi statistics of the electrons and the large disparity between the values of superfluid and Fermi velocity in these systems.

\end{abstract}
\maketitle

\section{Introduction}

Schr\"{o}dinger's cat paradox~\cite{Schroedinger35} emphasizes how basic quantum concepts such as superposition that are routinely applied and accepted in the microscopic description of matter appear to contradict basic human experience when 
augmented to a macroscopic scale. Quantum mechanics does not itself provide any intrinsic size limitation, i.e., it does not predict a critical number of particles at which such superpositions would be impossible. Yet there still remain questions as to whether there are emergent limitations at some length scale between the micro- and macroscopic, or even whether extraneous factors limit the size of superpositions. This question is fundamentally unresolved~\cite{Leggett2002}.   
Hints towards an answer would be given by experiments that realize truly macroscopic superpositions, which would provide evidence against macroscopic realism~\cite{Leggett85}.  
There are numerous
experiments attempting to produce such macroscopic 
superposition states in a variety of different physical systems~\cite{ArndtNairzVos-Andreae1999,
Haroche03,Haroche08,WalHaarWilhelm2000,Friedman00,HimeReichardtPlourde2006}.
Yet determining the actual size 
of these states (sometimes referred to as Schr\"odinger's 'cattiness'~\cite{Leggett2002}) is a non-trivial theoretical question that needs to be solved in tandem with the experimental realizations.  This includes three aspects.  First, one must ascertain or specify which types of degrees of freedom participate in the superposition, and which are irrelevant. 
For example,
trapped ions in a superposition state of internal degrees of freedom interact with the motional degrees of freedom in the ion trap, yet these would not be included in a superposition size measure for the internal states.  More generally, degrees of freedom that are not accessible in a particular experimental realization, e.g., because of too high energy, are not included. 
Thus in a low-temperature experiment one would not determine the superposition size in terms of quarks.
Second, once it is agreed which elementary constituents are involved, one needs to determine what is 
the actual number of particles or modes that participate in the superposition,
in the sense of being in distinguishably different states in the branches of the superposition.  The latter can be a highly non-trivial calculation for an interacting quantum system.  Third, it is important to then assess this number in the context of the observable 
and controllable physical parameters of the superposition state.

One attractive candidate system in  this program is the superconducting flux qubit.
Flux qubits are composed of superconducting loops of between one and many micrometres, that 
contain one or more Josephson
junctions. They can realize superpositions of states of a macroscopic electrodynamic variable, the circulating current and its concomitant magnetic flux, attributable to $10^6$-$10^{10}$ 
electrons~\cite{Leggett2002,WalHaarWilhelm2000,Friedman00,Mooij99}.
Superpositions
of the form $\leftcurrket + \rightcurrket$~\cite{WalHaarWilhelm2000,Friedman00} and coherent oscillations between these states have both been demonstrated~\cite{Chiorescu03,Insight}.
However, even though many particles are involved and the fluxes are macroscopically distinct in the two branches of the superposition, it does not necessarily follow that the actual {\em size} of the state, i.e., the number 
of particles that are in superposition
in the sense of being
in different states in the two branches, is macroscopic as well. Furthermore, this size can not be determined from the engineering of these experiments, for which the macroscopic electrodynamic
variables are sufficient~\cite{WalHaarWilhelm2000,Friedman00,Mooij99,Chiorescu03,Insight}.  
Finally, a microscopic calculation has to carefully take into account that the superconducting states is composed of (paired) electrons.  Because of their Fermion nature, these always occupy a finite 
volume in momentum space (unlike Bosons). Moreover, they are indistinguishable.  It is thus important to devise a measure that does not assume distinguishable electrons and that correctly takes both the indistinguishability and Fermion statistics of the electrons into account.
Theoretical estimates have come to vastly different proposals for this size, reflecting their different underlying assumptions:
i) counting all electrons of Cooper pairs in the supercurrent gives a size of 10$^6$-10$^{10}$~\cite{Leggett2002,WalHaarWilhelm2000,Friedman00,Leggett2005}, which is at least mesoscopic, if not approaching macroscopic sizes,
ii) analyzing the number of electrodynamic charge states within the macroscopic, circuit Hamiltonian
approach leads to a trivial effective size of 1-2~\cite{MarquardtAbelDelft2008}. 
However, neither of these approaches is fullly microscopic in the sense of characterizing the interacting electron system in the superconducting loop.

Here we place a bound on a broad range of microscopic measures of superposition size by asking the following direct question:
how many electrons behave distinguishably
differently in the two branches
$\leftcurrket$ and $\rightcurrket$?   We first outline the microscopic analysis of the superposition state that is necessary to estimate this quantity and then 
we present the bound for electrons in superpositions  of flux states.  This bound derives from an operational measure for superposition size that is determined by the degree of distinguishability of the two branches by an $n$-particle measurement~\cite{KorsbakkenWhaleyDubois2007}. 
We will show that for all experimentally realistic flux qubits, this bound results in an estimate of superposition state size that is considerably smaller than the number of electrons carrying the supercurrent but that is also significantly larger than the trivial size of 1-2 estimated from the macroscopic description.  Our microscopic analysis reveals that the 
Fermi
statistics play a critical role in reducing the number of electronic degrees of freedom that participate in 
the quantum superposition to well below 
simple estimates
of the number of current carrying electrons.  We further discuss the questions raised by such an estimate of size in terms of microscopic constituents, for a system whose macroscopic behavior can also be fully described by a single collective variable. 

\section{Microscopic analysis}

In this Section we outline the analysis performed to 
connect the full microscopic description of the system, starting from the Hamiltonian of Bardeen, Cooper and Schrieffer (BCS), to its macroscopic state and show how to compute microscopic quantities such as density matrices and correlation functions in that state.   We consider a flux loop containing a single tunnel junction and employ the 
functional integral formalism for superconducting tunnel junctions that was first developed in Ref.~\cite{EckernSchonAmbegaokar1984}.

\subsection{Hamiltonian}

The BCS Hamiltonian for the electrons in the bulk of the superconductor has the following form (in Gaussian cgs units):
 \begin{equation}
 \begin{split}
 H_{\text{BCS}} \, &= \, \sum_{\sigma} \int \diff^3 \spvec{r} \, \left \{ \psi_{\sigma}^{\dag} \, \frac{-\hbar^2}{2m} \left [ \nabla - \frac{ie}{\hbar c} \spvec{A} \right ]^2 \psi_{\sigma} \, + e \psi^{\dag}_{\sigma} \EMscalar \psi_{\sigma} \, - \, \frac{g}{2} \psi_{\sigma}^{\dag} \psi_{-\sigma}^{\dag} \psi_{-\sigma} \psi_{\sigma} \right \}
 \end{split}
 \end{equation}
 where $\psi_{\sigma}$ is the 
 electron field operator, $e$ is the (negative) electron charge, and the sum is over spin indices $\sigma = \pm 1/2$. $\EMscalar$ is the electromagnetic scalar potential 
 The 
 quartic term is the effective Cooper pairing electron-electron interaction, and includes both electric forces and phonon interactions.\footnote{This implies that the scalar potential $\EMscalar$ only includes potentials due to \emph{external} fields, not generated by electron dynamics in the bulk of the superconductor.}
The free electromagnetic field is governed by the following Hamiltonian:
 \begin{equation}
 \begin{split}
 H_{\text{EM}} \, &= \, \int \diff^3 \spvec{r} \, \frac{1}{8\pi} \left [ \spvec{E}_{\text{ind}}^2 + \spvec{h}_{\text{ind}}^2 \right ]
 \, = \, \int \diff^3 \spvec{r} \frac{1}{8\pi} \left [ \left ( -\frac{1}{c} \dot{\spvec{A}}_{\text{ind}} - \nabla \EMscalar_{\text{ind}} \right )^2 + \left ( \nabla \times \spvec{A}_{\text{ind}} \right )^2 \right ]
\end{split}
\label{eq:H_EM}
\end{equation}
where $\spvec{h}$ is the magnetic field and $\spvec{A}_{\text{ind}} = \spvec{A} - \spvec{A}_{\text{ext}}$. Only the fields $\spvec{E}_{\text{ind}} = \spvec{E} - \spvec{E}_{\text{ext}}$ and $\spvec{h}_{\text{ind}} = \spvec{h} - \spvec{h}_{\text{ext}}$ \emph{induced} by the electron dynamics are connected with the electron Hamiltonian, and therefore we subtract the external fields from the total field in $H_{\text{EM}}$.  Note that the integral in Eq.~(\ref{eq:H_EM}) is over all space.

An effective description can now be developed as follows:
\begin{enumerate}
\item We assume that there are no free electric fields in the bulk of the superconductor, so that the only terms in this region will be  $H_{\text{BCS}}$ and the term $\left ( \nabla \times \spvec{A}_{\text{ind}} \right )^2$ of $H_{\text{EM}}$.
\item Both close to and inside the Josephson junction, any induced electric fields across the junction and their interaction with the surface electrons can be described by an effective capacitive term $\frac{1}{2} C \Voltage^2$, where $\Voltage$ is the line integral of $-\spvec{E} = \frac{1}{c} \dot{\spvec{A}} + \nabla \EMscalar$ across the junction.
\item Outside the conductor, the energy of the magnetic field can be re-expressed as an inductive term, $\frac{1}{2L} \left ( \Phi - \extflux \right )^2$, where $\Phi$ is the flux enclosed by the superconducting loop.
\item The dynamics of electrons close to the junction due to tunneling across the junction can be expressed as an effective tunneling Hamiltonian,
\begin{equation}
H_{\text{T}} \, = \, \sum_{\sigma} \int \diff^3 \spvec{r} \int \diff^3 \spvec{r}' \, \psi_{\sigma}^{\dag}(\spvec{r}) \,  T_{\spvec{r} \spvec{r}'} \psi_{\sigma}(\spvec{r}')
\end{equation}
where $T_{\spvec{r} \spvec{r}'}$ is a tunneling amplitude which is non-zero only when $\spvec{r}$ and $\spvec{r}'$ are close to and on opposite sides of the junction.
\end{enumerate}

The effective Hamiltonian of the total system, bulk superconductor and tunneling junction, is then
\begin{equation}
H \, = \, H_{\text{BCS}} + H_{\text{T}} + \frac{1}{8\pi} \int \diff^3 r \, \left ( \nabla \times \spvec{A}_{\text{ind}} \right )^2 \, + \frac{1}{2L} \left ( \Phi - \Phi_{\text{ext}} \right )^2 + \frac{1}{2} CV^2
\label{eq:H_eff}
\end{equation}
where $H_{\text{T}}$ is only relevant close to a junction and the integral over the vector potential is now taken only over the interior of the superconductor.

\subsection{Expectation Values}
\label{ch:corr}

The electron field operators can be written in terms of creation and annihilation operators $\qop{c}_{\spvec{k}}^{\dag}$ and $\qop{c}_{\spvec{k},\sigma}$ for plane wave modes with wave vector $\spvec{k}$ as 
\begin{equation}
\qop{\psi}_{\sigma}(\spvec{r}) = \int\diff^3 k \, \qop{c}_{\spvec{k},\sigma} e^{-i\spvec{k}\cdot \spvec{r}}. 
\label{eq:FieldOperatorExpansion}
\end{equation}
Our goal will be to calculate arbitrary correlation functions 
in the grand-canonical ensemble 
between electron field operators.  
This will then also allow us to calculate expectation values of any operator $\qop{O}$ that is a function of electron creation or annihilation operators. We shall be specifically interested in mode occupation numbers, both for single electron modes and for Cooper pair modes.  The former are given by  2-point 1 mode correlation functions of the form
$\left \langle \qop{c}_{\spvec{k},\sigma}^{\dag} \qop{c}_{\spvec{k},\sigma} \right \rangle$ 
and the latter by 4-point 2 mode correlation functions of the form $\mean{\qop{c}_{\spvec{k},\uparrow}^{\dag} \qop{c}_{-\spvec{k},\downarrow}^{\dag} \qop{c}_{-\spvec{k},\downarrow} \qop{c}_{\spvec{k},\uparrow}}$.  To achieve these goals, we use the finite-temperature functional integral formalism with imaginary time-variable $\tau$ and 
electron variables anti-periodic in $\tau$, having period $\beta = 1/kT$~\cite{NegeleOrland1988} 
All 
expectation values calculated here will be restricted to the pure ground state by taking the limit $\beta \rightarrow \infty$.

The desired expectation values of an operator $\qop{O}$ 
are given by the functional integral expression~\cite{NegeleOrland1988}:
\begin{equation}
\mean{\qop{O}\left ( \qop{\psi}, \qop{\psi}^{\dag} \right )} \, = \, \int \FuncD{\psi} \, \FuncD{\spvec{A}} \, \FuncD{\EMscalar} \, O \left ( \psi, \psi^* \right ) \, e^{-S[\psi,\spvec{A},\EMscalar]/\hbar}
\end{equation}
with action functional $S[\psi]$ 
given by\footnote{Going to imaginary time means that an extra factor of $i$ is picked up by $\EMscalar$, which formally is the time-component of the four-vector $A^{\mu}$, so that the potential term of $H_{\text{BCS}}$ becomes $-ie \psi_{\sigma}^* \EMscalar \psi_{\sigma}$. }
\begin{equation}
S[\psi,\spvec{A},\EMscalar] = \int_{0}^{\hbar\beta} \diff \tau \, \left \{ \sum_{\sigma} \int\diff^3 \spvec{r} \, \left [ \psi_{\sigma}^*(\spvec{r},\tau) \hbar\partdiff{}{\tau} \psi_{\sigma}(\spvec{r},\tau) - \mu \psi_{\sigma}^*(\spvec{r},\tau) \psi_{\sigma}(\spvec{r},\tau) \right ] + H \right \}
\label{eq:EffectiveActionExpression}
\end{equation}
where $\mu$ is the chemical potential and $H$ is given by Eq.~(\ref{eq:H_eff}). 
We define the generating functional $\GenFunc{Z}{\xi}$~\cite{NegeleOrland1988,Korsbakken2008,KorsbakkenWilhelmWhaley2008}:
\begin{equation}
\GenFunc{Z}{\xi} \, = \, \int \FuncD{\psi} \, \FuncD{\spvec{A}} \, \FuncD{\EMscalar} \, e^{-S[\psi,\spvec{A},\EMscalar] + \sum_{\sigma} \int_0^{\hbar\beta} \frac{\diff \tau}{\taumeasure} \, \int \diff^3 r \, \left [ \xi_{\sigma}^*(\spvec{r},\tau) \psi(\spvec{r},\tau) + \xi_{\sigma}(\spvec{r},\tau) \psi_{\sigma}^*(\spvec{r},\tau) \right ]}
\label{eq:GeneratingFunctional1}
\end{equation}

One can now eliminate the quartic term in $H_{\text{BCS}}$ by introducing an extra integral over an auxiliary field and applying the Hubbard-Stratonovich transformation~\cite{Hubbard59,Stratonovich58,NegeleOrland1988}.
This leads to $S[\psi,\spvec{A},\EMscalar]$ in Eq.~(\ref{eq:EffectiveActionExpression}) being replaced by the effective action:
\begin{multline}
S[\psi,\spvec{A},\EMscalar,\Delta,\ephase] \, = \, \int_{0}^{\hbar\beta} \diff\tau \int \diff^3 r \sum_{\sigma} \left \{ \psi_{\sigma}^* \left [ \hbar \partdiff{}{\tau} - \frac{\hbar^2}{2m}\left ( \nabla - \frac{ie}{\hbar c} \, \spvec{A} \right )^2 - ie\EMscalar - \mu \right ] \psi_{\sigma} \,\right. \\
\left. + \, \frac{1}{2} 
\left ( \psi_{\sigma}^* \psi_{-\sigma}^* \Delta e^{i\ephase} + \Delta e^{-i\ephase} \, \psi_{-\sigma} \psi_{\sigma} \right ) \ \, + \, \frac{1}{g} \Delta^2 + H_T + H_{EM} \right \},
\label{eq:S_HS}
\end{multline}
where  the auxiliary field $\Delta e^{i\ephase}$, with $\Delta$ real-valued, has been introduced, and where the functional integral in Eq.~(\ref{eq:GeneratingFunctional1}) now also runs over all configurations of the fields $\Delta$ and $\ephase$.  

For convenience, it is usual to make a transformation $\psi_{\sigma} \rightarrow \psi_{\sigma} e^{i\phi/2}$ to remove the factors of $e^{\pm i \phi}$ from Eq.~(\ref{eq:S_HS}). The effect of this is that the operator $\nabla$ is replaced by $\nabla  - i \nabla \phi / 2$, which can then be combined with the vector potential $\spvec{A}$ to define the gauge-invariant superfluid velocity $\vsvec$:
\begin{equation}
\vsvec \equiv -\frac{\hbar}{2m} \left ( \nabla \ephase + \frac{2e}{\hbar c}\spvec{A} \right ).
\label{eq:v_s}
\end{equation}
Here $m$ the single electron mass, $e$ the (signed) single electron charge, and $\rho_e$ the
superfluid \textit{electron} density, i.e., the effective superconducting electron density~\cite{Schmidt97}.  
$\vsvec$ is equal to the mean velocity of
superconducting electrons in the system and is related to
the superconducting electron current density via $\spvec{j} = e \rho_e \vsvec$.
Note that the density of superconducting electrons, $\rho_e$, is twice the density of Cooper pairs, $\sfdens$ (commonly referrred to as the pair superfluid density), while $\vsvec$ is equivalent to the center of mass velocity of Cooper pairs.

Further introducing the Nambu space~\cite{Nambu1960} in which the electron field is expressed as the vector
\begin{equation}
\nambuvec{\Psi} \, = \, 
\begin{pmatrix}
\psi_{\uparrow} \\
\psi_{\downarrow}^*
\end{pmatrix}
\end{equation}
transforms the electronic part of the effective action into a proper quadratic form:
\begin{equation}
S_{\text{el}} \, = \, \int_0^{\hbar\beta} \diff\tau \int \diff^3 r \, \nambuvec{\Psi}^{\dag}(\tau,\spvec{r}) \nambumat{G}^{-1}(\tau,\spvec{r};\tau',\spvec{r}') \nambuvec{\Psi}(\tau',\spvec{r}'),
\label{eq:S_el}
\end{equation}
with
\begin{equation}
\nambumat{G}^{-1} = \nambumat{G}_{\text{bulk}}^{-1} + \nambumat{T}_{\spvec{r}\spvec{r}^{\prime}},
\end{equation}
where
\begin{equation}
\nambumat{G}_{\text{bulk}}^{-1} \, = \, \left [ \hbar\partdiff{}{\tau} + \left [ -\frac{\hbar^2}{2m} \left ( \nabla + \frac{im}{\hbar} \vsvec \right )^2 - \mu - ie\EMscalar \right ] \nambumat{\sigma}_z - \Delta \nambumat{\sigma}_x \right ] \delta(\spvec{r}-\spvec{r}') \delta(\tau-\tau')
\end{equation}
and the tunneling matrix $\nambumat{T}$ is given by
\begin{equation}
\nambumat{T}_{\spvec{r} \spvec{r}'} \, = \,
\begin{pmatrix}
T_{\spvec{r}\spvec{r}'} & 0 \\
0 & -T_{\spvec{r}\spvec{r}'}^*
\end{pmatrix}
\delta(\tau-\tau').
\label{eq:Tmatrix}
\end{equation}
The matrix $\nambumat{G}_{\text{bulk}}^{-1}$ is seen to be the inverse of the Green's function for the field
equations for the electron field in the bulk of the superconductor.

The generating functional Eq.~(\ref{eq:GeneratingFunctional1}) can be simplified by integrating over the electron field $\psi$, since  the electronic part of the action $S_{el}$, Eq.~(\ref{eq:S_el}), is now quadratic in $\psi$.  This results in a Gaussian integral which yields
\begin{equation}
\GenFunc{Z}{\xi} \, = \, \int \FuncD{\ephase}\, \, \FuncD{\spvec{A}} \, \FuncD{\EMscalar} \, \FuncD{\Delta} \,e^{\int_0^{\hbar\beta} \diff\tau' \, \diff \tau \int \diff^3 r \, \diff^3 r' \nambuvec{\xi}^{\dag} \nambumat{G} \nambuvec{\xi} - \frac{1}{2}CV^2 - \frac{1}{2L}\left ( \Phi - \Phi_{\text{ext}} \right )^2 + \trace \log \nambumat{G}^{-1}}.
\label{eq:GeneratingFunctionalAfterGaussianIntegration}
\end{equation}
The functional Eq.~(\ref{eq:GeneratingFunctionalAfterGaussianIntegration}) gets its largest contributions from those field configurations for which $S[\spvec{A},\EMscalar,\Delta,\ephase]$ has a maximum, i.e., where the functional derivatives of $S$ with respect to each of the fields is zero, i.e. a saddle point. 
For fields where the second derivative is very large, $S$ will get its \textit{only} significant contribution from the saddle point values, i.e., the fields will behave classically. As we described below, this turns out to be the case for the fields $\spvec{A}$, $\EMscalar$ and $\Delta$ but the second derivative of the effective action with respect to the phase $\ephase$ is not necessarily macroscopic.  Hence we cannot (at the moment) fix $\ephase$ to a single classical value. and this remains a quantum variable over which the functional integral is to be carried out.

We now summarize the results of the saddle point analysis for $\spvec{A}$, $\EMscalar$, and $\Delta$.  
For $\spvec{A}$ we find
\begin{equation}
\frac{c}{4\pi} \nabla \times \nabla \times \left ( \spvec{A} - \spvec{A}_{\text{ext}} \right ) \, = \, \spvec{j} \, = \, - 
\frac{e\rho_e \hbar}{2m} \left ( \nabla \ephase + \frac{2e}{\hbar c} \spvec{A} \right ) \, \equiv \, e \rho_e\vsvec.
\label{eq:current}
\end{equation}
The identification of the far left-hand side with the current density $\spvec{j}$ is done using Maxwell's equations (Amp\`{e}re's law).
For $\EMscalar$ we find
\begin{equation}
\hbar \partdiff{\ephase}{\tau} \, = \, 2 e \EMscalar,
\end{equation}
which we recognize as the AC~Josephson equation.
For $\Delta$, since in the situation we are interested in, the magnetic and electric fields will be much weaker than any critical values or energy scales and slowly varying, we may expand $\trace \log \nambumat{G}^{-1}$ perturbatively. 
This leads to the following saddle point equation:
\begin{equation}
\rho_F g \int_{-\omega_D}^{+\omega_D} \diff E \, \frac{1}{2\sqrt{E^2 + \Delta^2}} \, \tanh \left ( \frac{\beta}{2} \, \sqrt{E^2 + \Delta^2} \right ) \, = \, 1,
\end{equation}
where $\rho_F$ is the density of states at the Fermi surface, $\omega_D$ is the Debye energy, and $E$ denotes the difference in energy between a given energy level and the Fermi energy. 
Note that this is equivalent to the self-consistency equation for the energy gap $\Delta$ in BCS theory. 
As all currents, fields,  and temperatures in the system are far below their critical values, we can thus replace $\Delta$ with its standard BCS value.

We are left with $\ephase$ as the only non-classical variable. Expanding $\tr \log \nambuvec{G}^{-1}$ close to the junction perturbatively in $T_{\spvec{r}\spvec{r}'}$ leads to the further simplification
\begin{equation}
\GenFunc{Z}{\nambuvec{\xi}} \, = \, \int \FuncD{\ephase} \, e^{- \int_{0}^{\hbar\beta} \frac{\diff\tau}{\hbar} \left [ \frac{1}{2} C \dot{\Phi}^2 - E_J \, \cos \frac{2\pi \Phi}{\Phi_0} + \frac{1}{2L} \left ( \Phi - \Phi_{\text{ext}} \right )^2 + \int \diff^3 r \, \nambuvec{\xi}^{\dag} \nambumat{G} \nambuvec{\xi} \right ]},
\end{equation}
where the flux $\Phi$ is related to the phase variable $\ephase$ through the flux regular flux quantization relation
\begin{equation}
\Phi = \frac{\ephase}{2\pi} \, \Phi_0,
\end{equation}
and  the Josephson energy $E_J$ is an effective quantity whose form depends on the details of the tunneling matrix elements (a simplified form is given in Ref.~\cite{AmbegaokarBaratoff1963}).  

We now obtain correlation functions (expectation values of products of creation and annihilation operators) by taking time-ordered functional derivatives of $\GenFunc{Z}{\xi}$ with respect to $\nambuvec{\xi}$. This yields products of Green's functions which are
 linear combinations of products of the equal-time 
Gorkov Green's functions~\cite{AbrikosovGorkovDzyaloshinski1975,EckernSchonAmbegaokar1984}.
We note that, as derived also in Ref.~\cite{EckernSchonAmbegaokar1984},  the zero temperature limit of 
these Green's function products yields dynamics for the order parameter $\ephase$ that is equivalent to that of a particle in an effective potential $E_J \cos \frac{2\pi\Phi}{\Phi_0} - \frac{1}{2L} \left ( \Phi - \Phi_{\text{ext}} \right )^2$, which is precisely the description used in the macroscopic theory of flux qubits~\cite{Insight}.

The full Green's function, $\nambumat{G}$, is written as a sum of the zeroth order bulk Green's function and perturbative contributions from the superfluid flow and junction tunneling:
\begin{equation}
\nambumat{G} \, = \left ( \nambumat{G}_{\text{bulk}}^{-1} + \nambumat{T_{rr'}} \right )^{-1} 
=  \, \nambumat{G}_0 + \delta\nambumat{G}_{\vsvec} + \delta\nambumat{G}_T.
\label{eq:Gmat}
\end{equation}
In order to compare states with different current distributions, the Green's functions must be expressed in terms of velocity relative to a stationary laboratory frame, i.e., using modes that are eigenfunctions of the laboratory frame momentum operator 
\begin{equation}
\hbar \hat{\spvec{q}} \, \equiv \, -i\hbar \nabla + m \vsvec,
\label{eq:GIwavevecDef}
\end{equation}
rather than of $\hbar\hat{\spvec{k}}=-i\hbar \nabla$.
These bases can be viewed as related by a Doppler shift. Given that $|\vsvec|$ is 
smaller than 
the critical velocity, the energetics can be assumed not to change between frames~\cite{deGennes66}.
As we describe in Section~\ref{sec:DNtot}, the correction $\delta\nambumat{G}_{\vsvec}$ is evaluated by perturbatively expanding $\nambumat{G}_0(\spvec{q},\spvec{q}')$ in terms of the superfluid velocity $\vsvec$.  This yields~\cite{Korsbakken2008,KorsbakkenWilhelmWhaley2008} 
(see also discussion in Section~\ref{subsec:loop}):
\begin{equation}
\delta \nambumat{G}(\spvec{q},\spvec{q}') \, = \, - \FTmeasure \delta (\spvec{q} - \spvec{q}') \, \Delta \left ( \Delta \nambumat{\sigma}_z - \Ek{\spvec{q}} \nambumat{\sigma}_x \right ) \, \frac{\spvec{q} \cdot \vsvec}{2 \left ( \Ek{\spvec{q}}^2 + \Delta^2 \right )^{3/2}},
\label{eq:DeltaGbulk}
\end{equation}
where $\nambumat{\sigma}_x, \nambumat{\sigma}_z$ are Pauli spin matrices in the Nambu space.
The first order contribution to the Green's function from the tunneling matrix, $\delta\nambumat{G}_T$,  is obtained as~\cite{Korsbakken2008,KorsbakkenWilhelmWhaley2008} 
(see also discussion in Section~\ref{subsec:tunnel}):
\begin{equation}
\lim_{\beta\rightarrow\infty} \delta\FTnambumat{G}_T(\spvec{q}_L = \spvec{q}_R) \,
= \, \frac{\Delta}{4\hbar \EkDeltaSq{\spvec{q}}^3} \, T_{\spvec{k}\spvec{k}} \,
\begin{pmatrix}
- \Delta \left ( 1 - e^{-i\GIDphase} \right )  &  \Ek{\spvec{q}} \left ( 1 + e^{i\GIDphase} \right )  \\
\Ek{\spvec{q}} \left ( 1 + e^{-i\GIDphase} \right )  &  \Delta \left ( 1 - e^{i\GIDphase} \right )
\end{pmatrix}.
\label{eq:DeltaGTGaugeIndependentMomentumConserving}
\end{equation}
Here $\Omega_k \equiv \sqrt{E_k^2 + \Delta^2}$, where $E_k$ is the electron kinetic energy, 
$E_k = \frac{\hbar^2 q^2}{2m} -\fermi$, with $\fermi$ the Fermi energy, and 
$\GIDphase \, \equiv \, \Delta \ephase - \frac{2ie}{\hbar c} \int_{x_L}^{x_R} \spvec{A} \cdot \diff\spvec{r}$ is the gauge invariant phase difference across the tunnel junction.  

\section{Superposition size measures}

The size of a superposition state of a many-particle system is not a unique physical observable defined by its experimental 
measurement procedure, leaving many ways how to define 
superposition size.
A na\"{i}ve way to define such a size would be to compare the absolute value of the difference in some suitable physical observable between the two branches to some characteristic atomic scale for
 that observable. However, as pointed out in Refs.~\cite{Leggett2002,Leggett1980}, this approach of estimating an ''extensive difference'' between the branches is too simplistic.  For example, the superposition of a single neutron going two ways through an interferometer involves a huge difference in both center-of-mass position and angular momentum around the center of the interferometer compared to atomic 
scales, yet still involves only a single neutron and would clearly not be considered a macroscopic superposition state~\cite{Leggett2002}. 
Similarly, for superposition states of flow, e.g., in flux qubits, merely comparing the difference in magnetic moment $\Delta \mu$ between the branches $\leftcurrket$ and $\rightcurrket$ to the magnitude of the Bohr magneton $\mu_B$ can yield the same value $\Delta \mu = n \mu_B$ for very different $n$ particle 
states.  
For example, neglecting electron indistinguishability, a state in which $n$ electrons are in orthogonal states in the two branches, each having the same angular momentum difference $\Delta L = \hbar$, results in the same $\Delta \mu$ as a state in which a single high energy electron carries the equivalent total angular momentum difference $\Delta L = n \hbar$ while all other electrons behave identically in both branches. Only the former case involves $n$ electrons behaving distinguishably differently in the two branches and would be acceptable as a superposition state of size $n$: in the latter case only a single electron behaves differently and this is clearly a superposition state of size unity.  However, for electronic states of flow, we also have to take into account the fact that electrons are themselves indistinguishable, with Fermi statistics.  This further complicates the analysis, requiring an accounting of how many electronic {\em modes} behave differently in the two branches, i.e., have different occupation numbers, rather than how many {\em individual particles} behave differently.   This example suggests that in addition to differences between macroscopic variables, one needs to i) consider also the number of 
microscopic constituents that are behaving differently in the two branches when defining the effective size 
 of a many-body superposition state, and ii) take the fundamental particle statistics into account when defining these microscopic constituents. Since electrons satisfy Fermi statistics, the correct microscopic constituents for analysis of superposition states in flux qubits are modes of electron momentum, occupation of which can be different in the two branches. 

Several possible definitions have been put forward for a superposition size measure based on analysis of
microscopic degrees of freedom~\cite{MarquardtAbelDelft2008,Leggett1980,KorsbakkenWhaleyDubois2007}.  While all of these are hard
 to calculate for complex many-body states such as the electron states constituting a flux qubit, the distinguishability measure developed by two of us in Ref.~\cite{KorsbakkenWhaleyDubois2007} admits an upper bound that can be calculated using the microscopic functional integral formalism described above. The distinguishability measure is an operational measure that asks what is the largest number of subsets of elementary constitutents such that measuring all constituents in any one subset causes the superposition to collapse to one branch with some specified probability.  This problem is equivalent to that of determining the minimum number of microscopic constituents that have to be measured in order to distinguish the two branches to some specified precision. Ref.~\cite{KorsbakkenWhaleyDubois2007} showed that this leads to a superposition size of $N/n_{min}$, where $n_{min}$ is the first value for which the probability of successfully distinguishing the branches
is larger than $1-\delta$, where  
$\delta$ is the desired precision.
We adapt this here to consideration of the number of electron modes $n$ that need to be measured in the $N$ mode electronic system.   For simplicity we give the analysis here in terms of single electron modes.
It is important to note that the single electron modes combine to make Cooper pairs rather than excitations. 
As we show in more detail below, identical results are obtained if Cooper pair modes are used in our analysis.  However, we prefer to make the primary analysis with single electron modes since these 
highlight the underlying Fermi statistics, which as noted earlier are also 
reflected in the fact that the BCS state occupies a finite volume in momentum space.
For small differences in mode occupations
\begin{equation}
\qop{\rho}^{(n)}_A - \qop{\rho}^{(n)}_B \, \simeq \, \sum_{i=1}^n \left [ \bigotimes_{j \ne i}
\begin{pmatrix}
\modeNk{\spvec{q}_j}  &  0  \\
0  &  1 - \modeNk{\spvec{q}_j}
\end{pmatrix}
\right ]
\otimes
\left [
\begin{pmatrix}
\DmodeNk{\spvec{q}_i}  &  0  \\  0  &  0
\end{pmatrix}
+
\begin{pmatrix}
0  &  0  \\  0  &  -\DmodeNk{\spvec{q}_i}
\end{pmatrix}
\right ].
\label{eq:FirstOrderrhoABDifference}
\end{equation}
Averaging the probability of inferring the correct branch by measuring $n$ selected modes
 \begin{equation}
P_n \, = \, \frac{1}{2} + \frac{1}{4} \trace \left \| \qop{\rho}^{(n)}_A - \qop{\rho}^{(n)}_B \right \| \,
\simeq \, \frac{1}{2} + \frac{1}{2} \sum_{i=1}^n \left | \DmodeNk{\spvec{q}_i} \right |
\label{eq:PnModeDistinguishability}
\end{equation}
over all possible choices of $n$ out of $\NumModes$ modes leads to 
\begin{equation}
\overline{P}_n \, \simeq \,  \frac{1}{2} + \frac{n}{2\NumModes} \, \DNtot
\label{eq:PnModeDistinguishabilityEstimate}
\end{equation}
where the second term on the right hand side is an upper bound, derived without regard to multiple mode occupancy. From this a lower bound on  the limiting value of $n$ for success is extracted as $n = \NumModes / \DNtot$, resulting in an upper bound for the superposition size of $\DNtot$, i.e., the total difference in occupation number between the two branches. 
This estimate places an upper bound on any such microscopic measure of superposition size by asking how
many electrons on average are in a different mode in one branch
relative to the other branch.
Any measure that gives a larger number than this must be counting
electrons that are in identical states in the two branches and are
therefore not actively contributing to the superposition.
Since our measure is based on a measurement, it is desirable (although, just as in a gedanken experiment, not essential for theoretical evaluation of the superposition size)  that the assumed experiment is, in principle, feasible.  Thus it is essential that the indistinguishability of electrons be incorporated in the 
superposition size measure.  $\Delta N_{tot}$, the total change in occupation numbers of
all electron modes in the system is indeed the only meaningful
indicator of how many electrons are affected when passing from one
branch of the superposition to the other, when the indistinguishability of electrons is taken into account. 
We return to this in Section~\ref{sec:discussion} when discussing the role of Fermi statistics in determining the numerically obtained values for superposition states of flux qubits (Section~\ref{subsec:loop}).  

For our calculation based on the BCS theory~\cite{BardeenCooperSchrieffer1957}, as outlined above, the natural choice of single electron modes are momentum eigenstates. 
We now show that this choice of single-particle basis also maximizes the occupation number differences between the two branches,
$\delta n_{\spvec{q}} = \leftcurrbra \qop{c}_{\spvec{q},\sigma}^{\dag} \qop{c}_{\spvec{q},\sigma} \leftcurrket \, - \, \rightcurrbra \qop{c}_{\spvec{q},\sigma}^{\dag} \qop{c}_{\spvec{q},\sigma} \rightcurrket$, which ensures that $\DNtot$ is a mode-independent and hence true upper bound on the superposition size.
$\delta n_{\spvec{q}}$  can be viewed as the diagonal elements of a matrix $\mathbf{D}$ with elements $D_{\spvec{q}\sigma, \spvec{q}' \sigma'} = \leftcurrbra \qop{c}_{\spvec{q},\sigma}^{\dag} \qop{c}_{\spvec{q}',\sigma'} \leftcurrket \, - \, \rightcurrbra \qop{c}_{\spvec{q},\sigma}^{\dag} \qop{c}_{\spvec{q}',\sigma'} \rightcurrket$. 
$\DNtot = \sum_{\spvec{k}} \left | \delta n_{\spvec{k}} \right |$ is then the sum of the absolute values of the diagonal elements of $\mathbf{D}$, i.e., $\DNtot = \trace \left \| \mathbf{D} \right \|$. It can readily be shown that this number is maximal in the basis in which $\mathbf{D}$ is diagonal~\cite{Korsbakken2008,KorsbakkenWilhelmWhaley2008}.  
For a superconductor we have $\mean{\qop{c}_{\spvec{k},\sigma}^{\dag} \qop{c}_{\spvec{k}',\sigma'}} \propto \delta(\spvec{k}-\spvec{k}') \delta_{\sigma \sigma'}$, so $\mathbf{D}$ is diagonal in 
a momentum basis.  Therefore the occupation number differences must be evaluated in a momentum basis and not in a position basis, in order to correctly evaluate our upper bound to the superposition size.

Since we use BCS theory, the effect of Cooper pairing on the electron dynamics is automatically included, but to confirm this explicitly we shall also define two-electron Cooper pair modes and show that using these modes yields the same result 
for branch occupation number difference as that obtained using the single electron modes.
 
\section{The number of electrons changing modes between branches}
\label{sec:DNtot}

\subsection{Loop current contribution}
\label{subsec:loop}

Superposition states in flux qubits can be described 
 as superpositions of clockwise and counterclockwise net circulating currents of the
form $\frac{1}{\sqrt{2}} \left ( \leftcurrket + \rightcurrket \right)$. While these 
currents have intrinsic quantum fluctuations, these fluctuations do not affect the quantitative estimates made here which make use of the experimentally measured average currents (and hence implicitly an 
averaged superfluid phase parameter, since the latter is linearly related to the current).
The saddlepoint solution of our path integral equations described in Section \ref{ch:corr} is then characterized by the following assumptions.  For qubits with thickness much smaller than the penetration depth $\lambda$, we can take the flow to be approximately uniform in the
lateral dimension, i.e., the thickness in a quasi-planar geometry. 
The current flow is conveniently characterized by the complex phase $\ephase$ of the 
superconducting order parameter $\Delta$, and on the electromagnetic vector potential $\spvec{A}$, combined in the gauge-invariant combination that defines the \emph{superfluid velocity}, Eq.~(\ref{eq:v_s}). 
From the structure of the collective variables at the saddlepoint, the Green's functions and hence the occupation numbers are obtained 
by Fourier transforming and taking imaginary time-ordered functional derivatives of the generating functional $\GenFunc{Z}{\xi}$. We now summarize the main points of this calculation.  
Full details will be given elsewhere \cite{KorsbakkenWilhelmWhaley2008}.

Recognizing that as long as
both the externally applied and the internally generated magnetic
fields are weak  and the cooling is sufficiently
adiabatic to avoid vortex formation, the entire electron
system may be decomposed in terms of the single electron modes.  The 2-point, 1 mode correlation function is obtained as
\begin{subequations}
\begin{align}
\bra{\Psi} \qop{c}_{\spvec{k}\sigma}^{\dag} \qop{c}_{\spvec{k}\sigma} \ket{\Psi} \, &= \, \lim_{\tau \rightarrow 0^+} \mean{\ftrafo{\psi}_{\sigma}^*(\spvec{k},\tau) \ftrafo{\psi}_{\sigma}(\spvec{k},0)} \\
&= \, \lim_{\tau \rightarrow 0^+} 
\hFuncDeriv{}{\ftrafo{\xi}_{\sigma}(\spvec{k},\tau)} \, \hFuncDeriv{}{\ftrafo{\xi}_{\sigma}^*(\spvec{k},\tau')} \, \GenFunc{Z}{\nambuvec{\xi}} \\
&=  \, -\lim_{\tau \rightarrow 0^+}  \hbar G_{\sigma\sigma}(\spvec{k},\tau';\spvec{k},\tau),
\end{align}
\label{eq:OneElectron2PointOperatorProducts}
\end{subequations}
where  
$G_{\sigma\sigma}$ is a matrix element of $\nambumat{G}$.
As noted earlier, in order to compare properties of states with different current flow, in particular the two branches of a flux qubit superposition with left and right circulating currents, the correlation functions must be expressed in terms of the laboratory frame wave vector $\hat{\spvec{q}} \, \equiv \, -i \nabla + m \vsvec/\hbar$.  
The loop current contribution to the single mode correlation functions is then obtained as
\begin{equation}
\begin{split}
\bra{\Psi} \qop{c}_{\spvec{q}\sigma}^{\dag} \qop{c}_{\spvec{q}\sigma} \ket{\Psi} \, &= \,
- \hbar \lim_{\tau\rightarrow 0^+} G_{\sigma\sigma}(\spvec{q}-\frac{m\vsvec}{\hbar},0;\spvec{q}-\frac{m\vsvec}{\hbar},\tau).
\end{split}
\label{eq:QBranchOccupation1}
\end{equation}
Provided that the current in the superconductor is small compared to the critical current $I_{\rm c,bulk}$, a condition which is satisfied in the flux qubit 
experiments~\cite{Friedman00,WalHaarWilhelm2000,HimeReichardtPlourde2006}, the superfluid velocity $\vsvec$ will be a small perturbative quantity and calculations can be carried out to 
first order in $|\vsvec|/\vcrit = |\vsvec|\fermiv m/ \Delta$, where $\fermiv$ is the Fermi velocity.  
 Expanding 
 $G_{\sigma\sigma}$ leads then to 
 \begin{equation}
\begin{split}
\bra{\Psi} \qop{c}_{\spvec{q}\sigma}^{\dag} \qop{c}_{\spvec{q}\sigma} \ket{\Psi} \, &= \,
 \frac{1}{2} 
\left ( 1 - \frac{\Ek{q}}{\EkDeltaSq{q}} \right ) + \frac{1}{2} 
\frac{\Delta^2}{\EkDeltaSq{q}^3} \, \hbar \spvec{q} \cdot \mean{\vsvec},
\end{split}
\label{eq:QBranchOccupation2}
\end{equation}
where $\mean{\vsvec}$ is the mean superfluid velocity averaged over the quantum state of the system. 
The average occupation number of a single electron mode is given by 
$n_{\spvec{q}} =\mean{\qop{c}_{\spvec{q},\sigma}^{\dag} \qop{c}_{\spvec{q},\sigma}}$ ($n_{\spvec{q}}$ is independent of $\sigma$ for any realistic magnetic field strength).  Thus Eq.~(\ref{eq:QBranchOccupation2}) provides the contribution from $\nambumat{G}_0$ and $\delta\nambumat{G}_{\vsvec}$ to the mode occupation number. 
 The \emph{difference} in occupation number $\delta n_{\spvec{q}}$ of a mode $(\spvec{q},\sigma)$ 
 between the two circulating current superposition branches, to first order in $\vsvec$, is then obtained as
\begin{equation}
\delta n_{\spvec{q}} \, \equiv \, \leftcurrbra \qop{c}_{\spvec{q},\sigma}^{\dag} \qop{c}_{\spvec{q},\sigma} \leftcurrket \, - \, \leftcurrbra  \qop{c}_{\spvec{q},\sigma}^{\dag} \qop{c}_{\spvec{q},\sigma} \rightcurrket \, = \, \frac{\Delta^2}{2\EkDeltaSq{\spvec{q}}^3} \, \hbar \spvec{q} \cdot \Dmeanvs.
\label{eq:deltank}
\end{equation}
 
Before evaluating this difference for the recent flux qubit experiments, we summarize the corresponding analysis in terms of Cooper pair modes.  A Cooper pair mode accommodating two Cooper paired electrons with intrinsic momenta and spins $(\spvec{k},\uparrow)$ and $(-\spvec{k},\downarrow)$ is defined by the two-electron creation operator $\qop{C}_{\spvec{k}}^{\dag} \equiv \qop{c}_{\spvec{k},\uparrow}^{\dag} \qop{c}_{-\spvec{k},\downarrow}^{\dag}$. The Cooper pair mode correlation function    $\mean{\qop{C}_{\spvec{k}}^{\dag} \qop{C}_{\spvec{k}}}$ is equal to the  two-point correlation function for two single electron modes, $\mean{\qop{c}_{\spvec{k},\uparrow}^{\dag} \qop{c}_{-\spvec{k},\downarrow}^{\dag} \qop{c}_{-\spvec{k},\downarrow} \qop{c}_{\spvec{k},\uparrow}}$ and the branch occupation number difference for Cooper pair modes is equal to
\begin{equation}
\delta\mathcal{N}_{\spvec{k,-k}} = \, \leftcurrbra  \qop{c}_{\spvec{k}\uparrow}^{\dag} \qop{c}_{-\spvec{k}\downarrow}^{\dag} \qop{c}_{-\spvec{k}\downarrow} \qop{c}_{\spvec{k}\uparrow} \leftcurrket - \leftcurrbra \qop{c}_{\spvec{k}\uparrow}^{\dag} \qop{c}_{-\spvec{k}\downarrow}^{\dag} \qop{c}_{-\spvec{k}\downarrow} \qop{c}_{\spvec{k}\uparrow} \leftcurrket.
\end{equation}
 Analysis of the corresponding correlation functions in the laboratory frame is complicated by the fact that two laboratory frame velocity-modes $\spvec{q} = \spvec{k} + m\vsvec/\hbar$ and $\spvec{q}' = \spvec{k}' + m\vsvec/\hbar$ with opposite spins and with wave vectors $\spvec{k}$ and $\spvec{k}'$ relative to $\vsvec$ will be coupled by the Cooper pair coupling only if $\spvec{k}' = -\spvec{k}$.  Hence  $\spvec{q}' =  -\spvec{q} + 2m\vsvec/\hbar$ and so whether two modes are coupled or not will in principle depend on the superfluid velocity.  This can lead to a change in mode definitions between the two circulating 
 current states that complicates the calculation of mode occupation number difference.  However, since for the experimental systems $m\vsvec/\hbar$ is much smaller than the momentum-space difference between distinct modes, this mismatch can be treated perturbatively.  Detailed analysis shows that if the laboratory frame modes are not perfectly correlated, i.e., $\spvec{q}' \neq - \spvec{q} + 2m\vsvec/\hbar$ for the value of $\vsvec$ in either of the branches, there can be only a second-order dependence on $\vsvec$, while if they are perfectly correlated, i.e., $\spvec{q}' = - \spvec{q} + 2m\vsvec/\hbar$ in one branch, a first order contribution can result~\cite{Korsbakken2008,KorsbakkenWilhelmWhaley2008}.  
 Furthermore, the maximum value of this first order contribution is exactly the same as that obtained from the single electron mode occupation number difference, i.e., 
$\delta\mathcal{N}_{\spvec{q},\spvec{q}'} =  \delta {n_{\spvec{q}} \delta(\spvec{q} + \spvec{q}' - 2m\vsvec / \hbar)}$.  
Thus, as might have been expected, the two participating single electron modes are perfectly correlated.  More importantly for the present discussion, this result implies that analysis of the branch occupation number difference of Cooper pair (two-electron) modes yields exactly the same effective superposition size as that obtained from analysis of the branch occupation number difference of single electron modes. 

We now use the single electron mode analysis to derive an expression for the branch occupation number difference and hence an upper bound on the effective superposition size in terms of experimentally measured quantities.  We start by finding the total number of electrons changing mode in a local spatial region, 
which is obtained by summing Eq.~(\ref{eq:deltank}) over all modes~$\spvec{q}$:
\begin{equation}
\begin{split}
\DNdensity(\spvec{r}) \, &= \, 2\pi \int \, dq \, q^2 \modedensity{\spvec{q}} \frac{\Delta^2 \hbar q}{2\EkDeltaSq{q}^3} \int_{0}^{1}  \, \dcostheta \, \left | \dvsvec(\spvec{r}) \right | \, \cos \theta \\
&\equiv \, \pi \DNkern_1 \, \left | \dvsvec(\spvec{r}) \right |.
\end{split}
\label{eq:DNtotModeSum}
\end{equation}
Here $\theta$ is the angle between $\spvec{q}$ and $\dvsvec$, $\modedensity{\spvec{q}}$ is the (unknown) density of modes per unit volume and we have denoted the integral of $\Delta^2 \hbar q^3 / 2\EkDeltaSq{q}^3$ over all modes $\spvec{q}$ by  $\DNkern_1$. In addition, to avoid double-counting electrons, i.e., counting them both as they leave one mode and enter another, we only sum over the modes for which $\delta n_{\spvec{q}} > 0$.
Using the relation between current and velocity, we may derive a related expression for $\Dj(\spvec{r})$, the local current density:
\begin{equation}
\Dj(\spvec{r}) \, = \, e  \int \, d^3q \modedensity{q} |\delta n(\spvec{q})| \frac{\hbar \spvec{q}}{m}.
\end{equation}
From Eq.~(\ref{eq:deltank}) it is evident that components of $\spvec{q}$ perpendicular to $\dvsvec$ give zero contribution, so we can replace $\spvec{q}$ by its parallel component $q \cos \theta$ to arrive at
\begin{equation}
\begin{split}
\Dj(\spvec{r}) \, &= \, 2\pi e \int dq \, q^2 \modedensity{q} \frac{\Delta^2 \hbar q}{2\EkDeltaSq{q}^3} 
\frac{\hbar q}{m} \left | \dvsvec(\spvec{r}) \right | \,
\int_{-1}^{1} \, \dcostheta \cos^2 \theta \\
&= \, \frac{4 \pi}{3} e \DNkern_2  \left | \dvsvec(\spvec{r}) \right |,
\end{split}
\end{equation}
where $\DNkern_2$ denotes the integral of $\Delta^2 \hbar^2 q^4 / 2m\EkDeltaSq{q}^3$ over all modes $\spvec{q}$.
To evaluate the integrals $\DNkern_1$ and $\DNkern_2$, we note that the denominator in Eq.~(\ref{eq:deltank}) strongly suppresses modes away from the Fermi surface, i.e., $\delta n_{\spvec{q}}$ is non-negligible only for modes close to this, as expected for modes participating in Cooper pairing. $\modedensity{q}$ and $q$ may then be replaced by their values at the Fermi surface, $\rhoF$ and  
$\Fermiq{q} = \sqrt{2m\mu}/\hbar$, respectively, while the $q$ dependence of $\EkDeltaSq{q}$ must be maintained since this varies significantly over the range of $q$. The integrals then yield $\DNkern_1 =  \rhoF 2m^2\mu/\hbar^3$ and $\DNkern_2 =(\hbar \Fermiq{q} /m)\DNkern_1$, which allows the unknown $\modedensity{q}$ to be eliminated, to obtain the local occupation number difference in terms of the local current difference between the two branches:
\begin{equation}
\delta n(\spvec{r}) \, = \, \frac{3\, \left | \Dmeanjpos{\spvec{r}} \right |}{4 \, e \fermiv}.
\label{eq:dos}
\end{equation}
Finally, integrating over the entire volume of the superconductor yields the total branch occupation number difference $\DNtot$, the total number of electrons that are in different modes in the two branches, as
\begin{equation}
\DNtot \, = \, \frac{3\length}{4\, e \fermiv}
\, \DIp,
\end{equation}
where $\length$ is the total length of the main superconducting loop of the flux qubit and $\DIp$ is the
experimentally measured difference in ``persistent current''~\cite{WalHaarWilhelm2000} between the superposed branch states.  
We note that this derivation assumes that the current distribution is homogenous on scales less than or equal to the Fermi wavelength but not necessarily beyond this length scale.  Also, while the use of experimental persistent current values implicitly incorporates an average over quantum fluctuations in the superconducting phase parameter $\phi$,  explicit measurement of current fluctuations could allow a higher order analysis.

We have evaluated the occupation number difference bound on the effective superposition size for all  reported experimental demonstrations of flux state superpositions to date.  The relevant experimental parameters and corresponding values of $\DNtot$, the effective superposition size, 
are listed in Table~\ref{tbl:ExperimentsAndEffectiveCatSizes}
for
the three recent experiments~\cite{Friedman00,WalHaarWilhelm2000,HimeReichardtPlourde2006}.\begin{table}
\begin{tabular}{ | c | c | c | c | c | c | c |}
\hline
Experiment  &  Material  &  $\fermiv$  &  $L$  &  $\DIp$  &  $\Delta \mu$ &\DNtot  \\ \hline
SUNY  &  Nb  &  $1.37 \times 10^6 \un{m/s}$  &  $560 \un{\mu m}$  &  $2$--$3\un{\mu A}$  & $5.5-8.3 \times 10^9\mu_B$ & $3800$--$5750$\\
Delft  &  Al  &  $2.02 \times 10^6 \un{m/s}$  &  $20 \un{\mu m}$  &  $900\un{nA}$  & $2.4\times 10^6\mu_B$ & $42$  \\
Berkeley  &  Al  &  $2.02 \times 10^6 \un{m/s}$  &  $183 \un{\mu m}$  &  $292 \un{nA}$  & $4.23\times 10^{7}\mu_B$ & $124$  \\
\hline
\end{tabular}
\caption{Parameters and effective superposition sizes for current superposition states produced at SUNY~\cite{Friedman00}, Delft~\cite{WalHaarWilhelm2000} and Berkeley~\cite{HimeReichardtPlourde2006}.
$\fermiv$ is the Fermi velocity, $L$ the length of the superconducting loop, $\delta I_p$ the measured difference in persistent current between the two branches and $\delta \mu = A \delta I_p$ is the different in magnetic moment, where $A$ is the area enclosed by the loop.  $\DNtot$ is the effective superposition size, 
i.e., the total number of electrons participating in the superposition state.
}
\label{tbl:ExperimentsAndEffectiveCatSizes}
\end{table}
The largest numbers are found for the SUNY
experiment~\cite{Friedman00} carried out between excited states of a single-junction RF-SQUID configuration, while the Delft~\cite{WalHaarWilhelm2000} and Berkeley~\cite{HimeReichardtPlourde2006} experiments were both made with three-junction flux qubits that generated a superposition of degenerate ground states. The latter are very different in their geometric size. 
We also list in Table~\ref{tbl:ExperimentsAndEffectiveCatSizes} the corresponding values for the difference in the two macroscopic observables current and magnetic moment between the two branches,  respectively $\delta I_p$ and $\delta \mu=A\delta I_p$, where $A$ is the area enclosed by the superconducting loop.  Before discussing these results, we first briefly summarize the possible contributions from tunneling through the junction and the effect of scattering from impurities.

\subsection{Tunneling contribution}
\label{subsec:tunnel}

The estimates in Table~\ref{tbl:ExperimentsAndEffectiveCatSizes} include only the effect of different circulating currents in the two branches of the superposition.  This is the only contribution in the bulk of the superconductor, far from any junctions and where the tunneling contribution to the Green's function $\nambumat{G}$ is negligible.   However close to the junction, $\delta\nambumat{G}_T$ is not negligible and electron tunneling through the junction can then also contribute to the difference in mode occupation in the two branches of the superposition.  
Physically, this corresponds to the fact that close to the junction up to a depth of order $\xi_0=v_F/\Delta$, the Cooper pairs will be in superposition between both sides of the junction, rather than being described by rigid, directional superflow.
 This gives rise to an additional contribution to the superposition size when the modes are measured close to the junction. Detailed analysis of this using the form of $\delta\nambumat{G}_T$ given in Section~\ref{ch:corr} (i.e., neglecting the superfluid contribution here) leads to a mode occupation difference of~\cite{Korsbakken2008,KorsbakkenWilhelmWhaley2008}
\begin{equation}
\DmodeNk{\spvec{q}_L \spvec{q}_R}^{\pm} \, = \, \frac{\Delta^2}{\left ( \EkDeltaSq{L} + \EkDeltaSq{R} \right ) \EkDeltaSq{L} \EkDeltaSq{R}} \, \left | T_{\spvec{q}_L \spvec{q}_R} \right |.
\label{eq:CrossJunctionOccupationNumberChange}
\end{equation}
Evaluation of this quantity assuming momentum conservation during tunneling and using the parameters of the main tunnel junction in the Delft SQUID~\cite{WalHaarWilhelm2000} leads to the following estimate for the tunneling contribution to the occupation number difference close to the junction
\begin{equation}
33 \, \le \DNtot^{T} \, \le 43,
\label{eq:tunnelnumbers}
\end{equation}
which is the same order of magnitude as the number of electrons changing modes in the bulk of the flux loop for this qubit (second row in Table~\ref{tbl:ExperimentsAndEffectiveCatSizes}).  
Thus the tunneling dynamics of electrons close to the junction do not significantly alter the overall superposition size estimates made from microscopic consideration of the current carrying electrons in the bulk superconductor - both yield the same order of magnitude.
Note, that while these contributions are in general {\em not} simply additive,  if the range of tunneling is restricted to the vicinity of the junction, however, 
then adding them does provides a reasonable bound on the total superposition size that consistently includes both bulk and tunneling electron contributions.

\subsection{Dirty Superconductors}
\label{subsec:dirty}

The analysis above has addressed ballistic superconductors at $T=0$.  We now consider the effect of impurities and defects on these calculations of cat size in terms of the number of participating microscopic degrees of freedom.  
First, we note that while superconductors may in general be subject to inelastic effects from magnetic impurities, their concentration in modern nanofabricated samples is, however, extremely low, to the extent that there are no 
impurity-induced states in the gap, the density of 
states is perfectly BCS-like~\cite{Pothier97b}
and the dephasing times of flux qubits~\cite{Bertet05} are longer than expected in the presence of many magnetic impurities~\cite{Stamp03}.
However, the superconductor may still be 'dirty' by virtue of elastic scattering of the  Cooper pair electrons from impurities, which is likely due to the polycrystalline nature of an evaporated superconducting film. This results in typical diffusion constants of order 
$D=10^{-2}$ m$^2$/s, corresponding to a mean free path $\ell\simeq 10^{-8}$ m, which may be shorter than the 
coherence length $\xi_0 \sim \hbar\fermiv/\Delta$.  
 We now consider how this scattering affects the mode occupation number difference $\delta n_{\spvec{q}}$ in the dirty limit where  
 $\ell < \xi_0$ ($\xi_0 = 1.6 \times 10^{-6}$ m in Al and $3.8 \times 10^{-8}$ m in Nb). 

We need to analyze how the Green's function $\nambumat{G}(\spvec{q},\spvec{q}')$ is modified by elastic scattering from impurities.  We shall restrict our discussion here to the bulk contribution, $\nambumat{G}_{\text{bulk}} = \nambumat{G}_0 + \delta\nambumat{G}_{\vsvec}$ and not consider the tunneling contribution.
Tunneling influences the Green's functions over a length corresponding to the 
appropriate dirty-limit coherence length $\xi_{0,D}=\sqrt{\ell \xi_0/3}$~\cite{Belzig99}, which is necessarily smaller than $\xi_0$ in this limit. Thus, we expect a smaller tunneling contribution than in the ballistic case.
For weak scattering, the modified bulk Green's function can be analyzed using the Dyson expansion
\begin{equation}
\nambumat{\tilde{G}} \, =  \nambumat{G}_{\text{bulk}} + \nambumat{G}_{\text{bulk}}U\nambumat{G}_{\text{bulk}}
\label{eq:Dyson}
\end{equation}
where $U$ is the electron-impurity interaction.  Expanding as before in the small parameter  $|\vsvec|/\vcrit = |\vsvec|\fermiv m/ \Delta$ leads to a zeroth order term $\nambumat{G}_0 U \nambumat{G}_0$ and a first order term $\nambumat{G}_0 U \delta\nambumat{G}_{\vsvec} + \delta\nambumat{G}_{\vsvec}  U \nambumat{G}_0$.  We take the electron-impurity interaction $U$ to be defined by
\begin{equation}
U(\vec{r}) = \sum_j U_0 \delta(\vec{r} - \vec{r}_j)
\label{eq:impurity_r}
\end{equation}
where the sum goes over all impurities in the superconducting loop.
The modified Green's functions may then be evaluated using the explicit solution for the zeroth order Green's function,
\begin{equation}
\nambumat{G}_0(\spvec{k},\tau;\spvec{k}',\tau') \, = \, \FTmeasure
\delta(\spvec{k}-\spvec{k}') \, \frac{e^{-\frac{\EkDeltaSq{k} |\tau-\tau'|}{\hbar}}}{2\hbar} \\ 
\times
\begin{cases}
\left [ \identity + \frac{1}{\EkDeltaSq{k}} \left ( \Ek{k} \nambumat{\sigma}_z + \Delta \nambumat{\sigma}_x \right ) \right ] \, 
& 0 < \tau-\tau' < +\frac{1}{2}\hbar\beta \\
(-1) \left [ \identity - \frac{1}{\EkDeltaSq{k}} \left ( \Ek{k} \nambumat{\sigma}_z + \Delta \nambumat{\sigma}_x \right ) \right ] 
. & -\frac{1}{2} \hbar\beta < \tau-\tau' < 0
\end{cases}
\label{eq:FreeGreensFunctionResultk}
\end{equation}
Carrying out the time and momentum space integrals, we find that the zeroth order term $\nambumat{G}_0 U \nambumat{G}_0$ is equal to zero and that the two contributions to the first order term cancel.  Thus, to first order in the superfluid velocity 
and the electron-impurity interaction, elastic scattering from impurities has no effect on the single electron Green's function and hence no effect on our estimate of the effective superposition size.  

While higher order terms in the Dyson expansion may also give rise to contributions that are first order in $\vsvec$, a more complete analysis based on quasiclassical Green's function~\cite{Rammer86,Belzig99,Anthore03}
 shows that these conclusions are independent of the impurity concentration as long as $\ell\gg \lambda_F$, where $\lambda_F$ is the Fermi wavelength. The latter can 
always be assumed, because $\ell \simeq \lambda_F$ would put the material close to the Anderson metal-insulator transition in the normal state~\cite{Anderson58}, hence the superconductor would not be a 'good metal' any more.  
The analysis proceeds as follows. In the dirty limit, the superfluid velocity is related to the phase gradient by 
$\vsvec = D \left ( \nabla \ephase + \frac{2e}{\hbar c}\spvec{A} \right )$,
where $D=\frac{1}{3} v_F\ell$ is the diffusion constant~\cite{Anthore03,Anthore03b}. The supercurrent density
is still proportional to $\vsvec$ and is found to be independent of $D$~\cite{Anthore03}.  Furthermore, at a given current, the gauge-invariant phase gradient 
$\nabla\phi+\frac{2e}{\hbar}\vec{A}$ is  proportional to $1/D$. Now the angular dependence of the Green's function in the dirty limit is 
only weakly anisotropic, with a large $s$-wave component that does not depend on the direction of the circulating current (and hence does not contribute to the difference
in mode occupation)
and a small $p$-wave component that is proportional to 
$\ell \left(\nabla\phi+\frac{2e}{\hbar}\vec{A}\right)$~\cite{Belzig99,Rammer86}. Using the above-mentioned dependence of the gauge invariant phase gradient on the diffusion coefficient, we then find that at a given current, the $p$-wave component of the Green's function is independent of the mean free path $\ell$.  Futhermore, for the situation of a 
homogeneous circulating current, the angular dependence of the Green's function is restricted to $s$- and $p$-wave components only~\cite{Rammer86}.  Thus the components of the Green's function relevant for evaluation of $\delta n_q$ in superconducting flux loops are independent of the impurity concentration and hence our estimates of effective superposition size apply also to the dirty superconductor regime.

\section{Discussion}
\label{sec:discussion}

The results in Table~\ref{tbl:ExperimentsAndEffectiveCatSizes} thus show that while not trivially small, the effective superposition sizes 
as bounded by the number of electrons that are in different modes in the two branches of the
 superposition are considerably smaller than 
previous estimates of the
number of electrons carrying the supercurrent, which were based on simply counting all electrons within a London penetration depth of the surface~\cite{Friedman00,WalHaarWilhelm2000,Leggett2002,Leggett2005}.  
In fact these estimates can now be replaced by the number of electrons in different modes within the two branches, $\Delta N_{tot}$, since 
only this number of electrons is actually moving in opposite directions in the two branches. Specifically,
our microscopic analysis of all electronic degrees of freedom shows that when electron indistinguishability is taken into account, the number of electrons
responsible for the observed difference in supercurrent is actually far smaller than the number of electrons within the London penetration depth.
  Furthermore, there is a marked contrast between the fact that the values of $\Delta N$ are mesoscopic,
while the two branches nevertheless have macroscopically distinct values of the two observables, persistent current and magnetic moment.  We show below that this discrepancy is due to the Fermi statistics of the electrons.

While the sizes obtained here are generally larger than typical estimates of 
size for superposition
states realized in molecular and optical systems~\cite{ArndtNairzVos-Andreae1999,Haroche08} and could reasonably be termed mesoscopic,
they are still well
short of anything that could be considered truly macroscopic. We therefore conclude that the superposition states in flux qubits
are superpositions 
of relatively small numbers of particles 
that nevertheless result in a
large difference in magnetic moment and current between the two branches 
because
the electrons are circulating in opposite directions at high speeds (the Fermi
velocity, which is of order $10^6$ms$^{-1}$) around a path enclosing a relatively large surface area. The actual number of electrons that would be found to be behaving
significantly differently in the two branches if one could measure them at the microscopic
level is, however, seen to be quite modest.
This situation is illustrated 
qualitatively in Fig.~1 by 
schematics of the shells at the Fermi surface that contain the occupied one-electron levels participating in Cooper pairing within each of the two branches of the superposition. The effect of 
realizing current flows $\pm\vsvec$ in the two branches is to shift each such Fermi shell by a distance $\pm m\vsvec/\hbar$ in $\spvec{k}$-space. Compared to $m\fermiv/\hbar$, the radius of the Fermi sphere, this distance is extremely small, so that the two Fermi shells overlap strongly. The vast majority of modes therefore have identical occupation numbers in both branches: only modes in the non-overlapping regions in Fig.~1 have a change in occupation number. However, these regions are on \emph{opposite} sides of the Fermi sphere. 
Furthermore, simply stating that the two branches are separated by a slight shift of the entire Fermi sphere is
incompatible with the fact that electrons are indistinguishable. 
Because of the electron indistinguishability and Fermi statistics, rather than shifting a large number of electrons by the small distance $m\vsvec/\hbar$, the shift is instead properly described as a relatively small number of electrons being moved all the way from one side of the sphere to the other and thereby changing their momentum by 
$\simeq 2k_F$.
This is an extremely large velocity change because of the large diameter of the Fermi sphere, even though the small value of 
$\vs$ means that the volume of the non-overlapping regions and therefore the number of electrons moved are both small. This discrepancy between a large difference in the value of an observable quantity 
(current $\delta I_p$ or magnetic moment $\delta \mu$)
and the small number of particles actively involved in the superposition
by {\em distinguishably} changing modes, $\Delta N_{tot}$, thus derives from the fermionic character of electrons. 

The corresponding gap between difference in observable quantities and number of particles involved might be smaller for large scale superposition states in a bosonic system. Finding a physical  system to realize this constitutes a challenge for future study, both experimental and theoretical.
\begin{figure}
\includegraphics{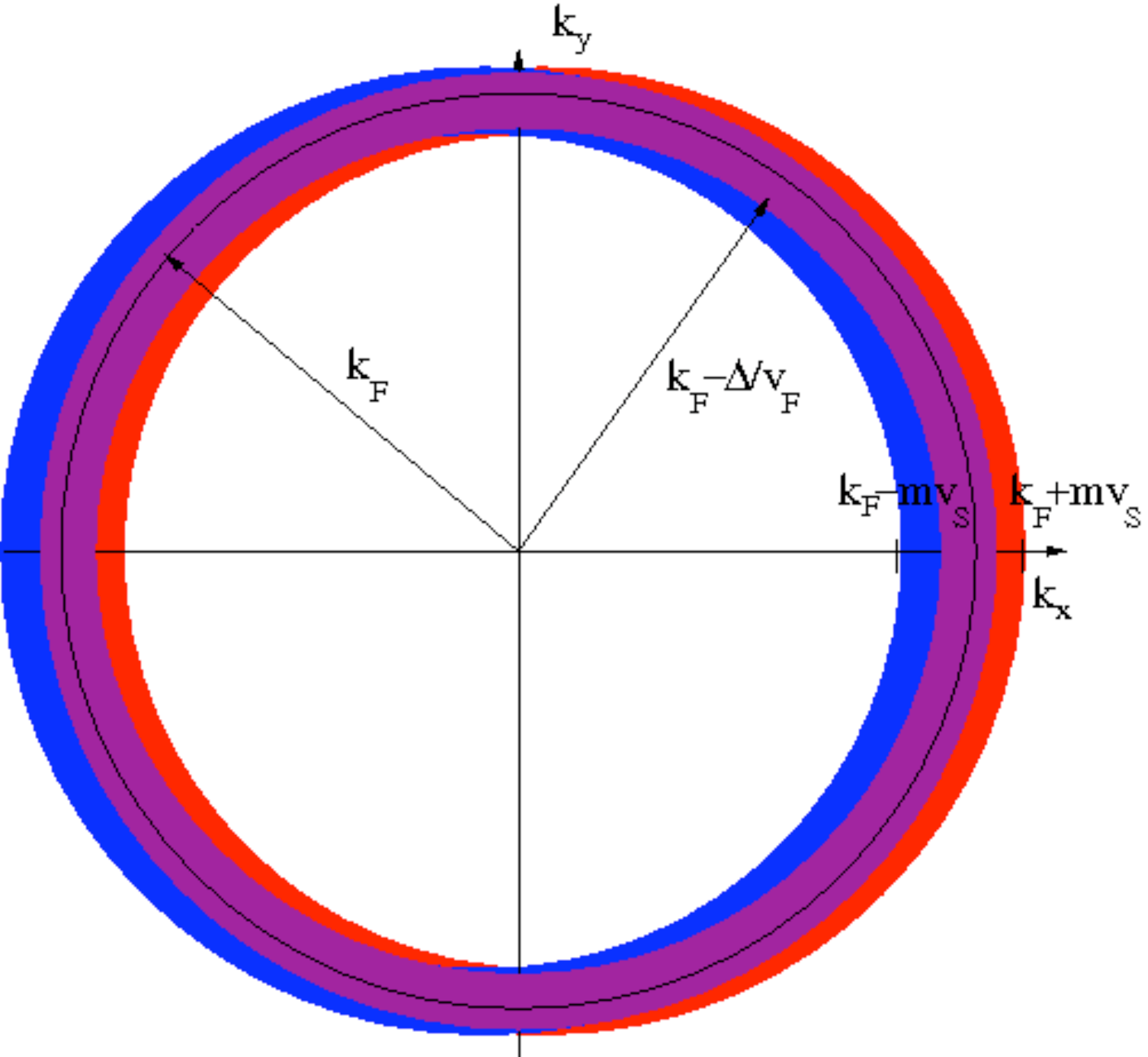}
\caption{Schematic of the two shells of single electron states at the Fermi surface involved in Cooper pairing for each of the two branches of the superposition $\leftcurrket + \rightcurrket$.  Each Fermi sphere is shifted by a distance $\pm 2m\vsvec/\hbar$ in $\spvec{k}$-space and the shells are color coded as red and blue, with the overlapping region denoted by the color sum, purple.  Note that the figure is \emph{not} to scale, with the shift being exaggerated relative to the diameter of the Fermi spheres in order to be visible. Only modes in the non-overlapping regions (red and blue) have different occupation numbers in the two branches.}
\label{fig:ShiftedFermiSpheres}
\end{figure}

\begin{figure}
\includegraphics[width=0.7\columnwidth]{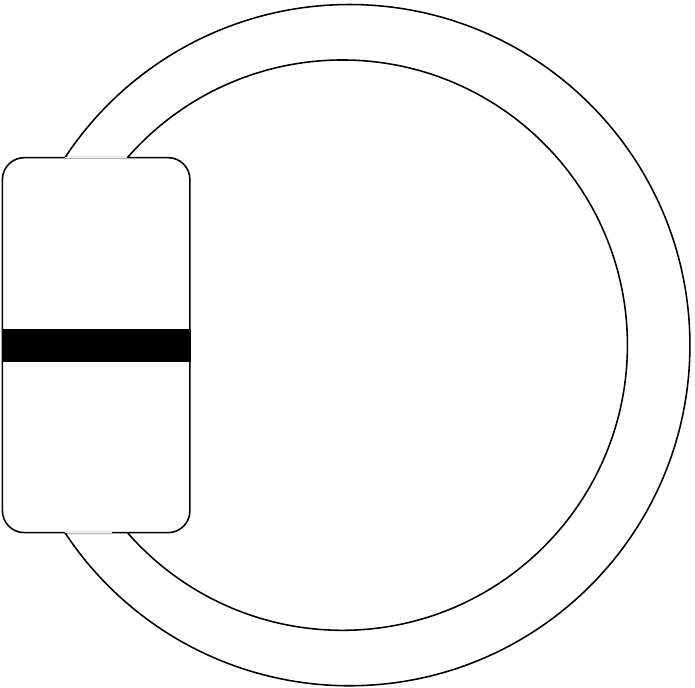}
\caption{\label{fig:cross} A schematic of a large Josephson junction corrected to a narrow wire, maximizing the change in kinetic energy of the loop electrons and thus the ratio of $\Delta N/N$.}
\end{figure}

From Eq.~(\ref{eq:deltank}) the maximum difference in occupation
number for any one mode is bounded by $\hbar \left | \spvec{q} \cdot
  \Dmeanvs \right | / 2\Delta$.  This number is always small unless current
differences are close to the critical current (more specifically, to the depairing current of the material), which is
impossible in a system where the current passes through Josephson
junctions. In fact, the ratio between critical and Josephson currents can be written as
$I_{\rm c, bulk}/{I_{\rm c, J}}=4\pi f R_N/{R_S}$,
where $f$ is the superconducting condensate fraction ($f=1$ at $T=0$), $R_N$ the junction resistance
in the normal state  and $R_S$ the Sharvin resistance of the
superconducting material~\cite{Schmidt97}.  
Given that Josephson junctions have high resistivity while
superconducting metals have low resistivity, this ratio is usually large.  
One strategy
to reduce it is to use large area Josephson junctions to obtain small $R_N$, 
connected by
superconducting wires with small cross-section to maximize $R_S$, as shown in 
figure \ref{fig:cross}.  
Another option is to increase the number of modes available, by scaling up the physical dimensions of the system. However, the number of modes scales only linearly with system size. To reach truly 
macroscopic superposition states in this way would therefore require the physical dimensions of the flux qubits to be scaled up by many orders of magnitude.   
While this would
certainly make it extremely
challenging to maintain or observe any superposition behaviour in the presence of environmental noise,
recent estimates indicate that coherence may still be observable in high inductance single Josephson junction loops with linear dimension of order 1 cm~\cite{Mooij09}.

The results presented above show that the microscopic number of effective electron constituents in flux qubit superpositions with dimensions accessible to experiments today, such that these constituents behave differently in the two superposition branches, is of order $10^2 - 10^3$.  
This addresses a long-standing question related to macroscopic quantum coherence~\cite{Leggett2002,Leggett1980}.
We  note that while this value derives from an operational measure that is independent of the physical system, asking only how many $n$-particle measurements are required to distinguish the two superposition branches, it is nevertheless useful to ask the further question of how these $n$-particle measurements may be realized and whether they may be decomposed into single particle measurements (for an example of the latter, see~\cite{KorsbakkenWhaleyDubois2007}).  Similarly, one may ask whether it is possible to control the $\DNtot$ constituent degrees of freedom individually, as is assumed in most applications of a GHZ state,
i.e., does the size of the superposition state reflect the number of useful dimensions (analogous to the distinction between useful and non-useful or ''fluffy bunny" entanglement~\cite{DunninghamRauBurnett2005})? 
Here we note that due to the large superconducting gap, actual realization of the distinguishability measure would require an $n$-electron Quantum Non-Demolition (QND) measurement that does not involve excitation across the gap.  
Such measurements have not been devised so far and would 
require bridging about one order of magnitude in energy between the qubit tunnel matrix elements and the energy gap.  Nevertheless, the present bound on this measure does provide a microscopic analysis 
of the number of electronic constituents that could behave in a Schr\"{o}dinger cat-like fashion, i.e., show (macroscopically) demonstrably different behavior in the two branches.  It is then a further question as to whether and how this demonstrably different behavior of the individual constituents may be realized.

\section{Conclusion}

We have used the functional path integral formalism to connect the microscopic and macroscopic description of flux qubits and to derive expressions for correlation functions of creation and annihilation operators.   This connection was then used to characterize the effective size of superposition states in these systems by evaluating the change in occupation number of electronic modes between the two branches of a flux superposition state 
that is characterized by macroscopic branch differences in physical observables.   We showed that this quantity constitutes an upper bound on the distinguishability measure developed previously by two of us.  
The results obtained here for flux qubits of physical dimension accessible today show that the number of electrons, or equivalently, of Cooper pairs, that participate actively in the superposition behavior is of order $10^2 - 10^3$, considerably less than the total number of electrons participating in the supercurrent.  
This result shows that even if there is no intrinsic size or number scale limiting the existence of macroscopic quantum superpositions, the quantum statistics of the constituent particles can nevertheless
be important for evaluating
the effective number of particles participating in a superposition whose branches are characterized by macroscopically distinct observables.  
In particular, in the case of flux qubits, our analysis has revealed that it is a combination of the indistinguishability and Fermionic character of electrons, together with the large numerical value of the Fermi
velocities, that is responsible for the large change in current and magnetic moment per particle between 
branches, rather than a change in state of a macroscopic number of electrons.
In summary, we see that a full microscopic treatment of the electrons in flux qubits shows that the superconducting flux qubit experiments performed to date 
provide neither a verification nor proof for the formation of true quantum superpositions on a scale beyond a few thousand 
microscopic constituent particles. The experimental quest for
superpositions on the truly macroscopic scale as well as the verification or
falsification of macrorealism by this route therefore remain open.

\ack
 We acknowledge J.I. Cirac and J. von Delft for early
   collaborations on related projects 
   as well as A.J. Leggett for stimulating discussions. This work was supported by NSF through the ITR program and by NSERC through the
   Discovery Grants program.


\section*{References}
\begin{thebibliography}{10}

\bibitem{Schroedinger35}
E.~Schr\"odinger.
\newblock Die gegenw\"artige {Situation} in der {Quantenmechanik}.
\newblock {\em Naturwissenschaften}, 23:807, 823, 844, 1935.

\bibitem{Leggett2002}
A.~J. Leggett.
\newblock Testing the limits of quantum mechanics: motivation, state of play,
  prospects.
\newblock {\em J.~Phys.---Condens.~Matter}, 14(15):R415--R451, Apr 2002.

\bibitem{Leggett85}
A.~J. Leggett and Anupam Garg.
\newblock Quantum mechanics versus macroscopic realism: Is the flux there when
  nobody looks?
\newblock {\em Phys. Rev. Lett.}, 54(9):857--860, 1985.

\bibitem{ArndtNairzVos-Andreae1999}
M.~Arndt, O.~Nairz, J.~Vos-Andreae, C.~Keller, G.~van~der Zouw, and
  A.~Zeilinger.
\newblock Wave-particle duality of c60 molecules.
\newblock {\em Nature}, 401(6754):680--682, 1999.

\bibitem{Haroche03}
S.~Haroche.
\newblock Quantum information in cavity quantum electrodynamics: logical gates
  entanglement engineering and "Schr\"{o}dinger-cat states".
\newblock {\em Phil. Trans. R. Soc. Lond. A}, 361:1339--1347, 2003.

\bibitem{Haroche08}
S.~Haroche, M.~Brune, and J.-M. Raimond.
\newblock Schr\"{o}dinger cat states and decoherence studies in cavity qed.
\newblock {\em Eur. Phys. J. Special Topics}, 159:19--26, 2008.

\bibitem{WalHaarWilhelm2000}
C.~H. van~der Wal, A.~C.~J. ter Haar, F.~K. Wilhelm, R.~N. Schouten, C.~J.
  P.~M. Harmans, T.~P. Orlando, S.~Lloyd, and J.~E. Mooij.
\newblock Quantum superposition of macroscopic persistent-current states.
\newblock {\em Science}, 290(5492):773--777, Oct 2000.

\bibitem{Friedman00}
J.R. Friedman, V.~Patel, W.~Chen, S.K. Tolpygo, and J.E. Lukens.
\newblock Quantum superposition of distinct macroscopic states.
\newblock {\em Nature}, 46:43, 2000.

\bibitem{HimeReichardtPlourde2006}
T.~Hime, P.~A. Reichardt, B.~L.~T. Plourde, T.~L. Robertson, C.-E. Wu, A.~V.
  Ustinov, and John Clarke.
\newblock Solid-state qubits with current-controlled coupling.
\newblock {\em Science}, 314(5804):1427--1429, 2006.

\bibitem{Mooij99}
J.E. Mooij, T.P. Orlando, L.~Levitov, L.~Tian, C.H. van~der Wal, and S.~Lloyd.
\newblock {Josephson} persistent current qubit.
\newblock {\em Science}, 285:1036, 1999.

\bibitem{Chiorescu03}
I.~Chiorescu, Y.~Nakamura, C.J.P.M. Harmans, and J.E. Mooij.
\newblock Coherent quantum dynamics of a superconducting flux qubit.
\newblock {\em Science}, 299:1869, 2003.

\bibitem{Insight}
J.~Clarke and F.K. Wilhelm.
\newblock Superconducting qubits.
\newblock {\em Nature}, 453:1031, 2008.

\bibitem{Leggett2005}
A.~J. Leggett.
\newblock The quantum measurement problem.
\newblock {\em Science}, 307:871--872, 2005.

\bibitem{MarquardtAbelDelft2008}
F.~Marquardt, B.~Abel, and J.~von Delft.
\newblock Measuring the size of a quantum superposition of many-body states.
\newblock {\em Phys. Rev. A}, 78(1):012109, 2008.

\bibitem{KorsbakkenWhaleyDubois2007}
J.~I. Korsbakken, K.~B. Whaley, J.~Dubois, and J.~I. Cirac.
\newblock Measurement-based measure of the size of macroscopic quantum
  superpositions.
\newblock {\em Phys. Rev. A}, 75(4):042106, 2007.

\bibitem{EckernSchonAmbegaokar1984}
U.~Eckern, G.~Sch\"{o}n, and V.~Ambegaokar.
\newblock Quantum dynamics of a superconducting tunnel junction.
\newblock {\em Phys. Rev. B}, 30(11):6419--6431, Dec 1984.

\bibitem{NegeleOrland1988}
J.~W. Negele and H.~Orland.
\newblock {\em Quantum many-particle systems}.
\newblock Frontiers in Physics. Addison-Wesley Pub. Co., 1988.

\bibitem{Korsbakken2008}
J.~I. Korsbakken.
\newblock {\em Schr\"{o}dinger's Lion or Schr\"{o}dinger's Kitten? --- Gauging
  the Size of Large Quantum Superposition States}.
\newblock PhD thesis, University of California, Berkeley, Berkeley, CA 94720,
  U.S.A., Fall 2008.

\bibitem{KorsbakkenWilhelmWhaley2008}
J.~I. Korsbakken, F.~K. Wilhelm, and K.~B. Whaley.
\newblock To be published.

\bibitem{Hubbard59}
J.~Hubbard.
\newblock Calculation of partition functions.
\newblock {\em Phys. Rev. Lett.}, 3:77, 1959.

\bibitem{Stratonovich58}
R.L. Stratonovich.
\newblock On a method of calculating quantum distribution functions.
\newblock {\em Sov. Phys. Doklady}, 2:416, 1958.

\bibitem{Schmidt97}
V.V. Schmidt.
\newblock {\em The physics of superconductors}.
\newblock Springer, Berlin, 1997.

\bibitem{Nambu1960}
Y.~Nambu.
\newblock Quasi-particles and gauge invariance in the theory of
  superconductivity.
\newblock {\em Phys. Rev.}, 117(3):648--663, Feb 1960.

\bibitem{AmbegaokarBaratoff1963}
V.~Ambegaokar and A.~Baratoff.
\newblock Tunneling between superconductors.
\newblock {\em Phys. Rev. Lett.}, 10(11):486--489, Jun 1963.

\bibitem{AbrikosovGorkovDzyaloshinski1975}
A.~A. Abrikosov, L.~P. Gorkov, and I.~E. Dzyaloshinski.
\newblock {\em Methods of quantum field theory in statistical physics},
  chapter~34.
\newblock Dover, New York, 1975.

\bibitem{deGennes66}
P.G. de~Gennes.
\newblock {\em Superconductivity of metals and alloys}.
\newblock Benjamin, N.Y., 1966.

\bibitem{Leggett1980}
A.~J. Leggett.
\newblock Macroscopic quantum-systems and the quantum-theory of measurement.
\newblock {\em Prog. Theor. Phys. Supp.}, 69:80--100, 1980.

\bibitem{BardeenCooperSchrieffer1957}
J.~Bardeen, L.~N. Cooper, and J.~R. Schrieffer.
\newblock Theory of superconductivity.
\newblock {\em Phys. Rev.}, 108(5):1175--1204, Dec 1957.

\bibitem{Pothier97b}
H.~Pothier, S.~Gueron, N.O. Birge, and D.~Esteve.
\newblock Energy distribution of electrons in an out-of-equilibrium metallic
  wire.
\newblock {\em Z. Phys. B}, 103:313--318, 1997.

\bibitem{Bertet05}
P.~Bertet, I.~Chiorescu, G.~Burkard, K.~Semba, C.J.P.M. Harmans, D.P.
  DiVincenzo, and J.E. Mooij.
\newblock Relaxation and dephasing in a flux-qubit.
\newblock {\em Phys. Rev. Lett.}, 95:257002, 2005.

\bibitem{Stamp03}
P.~C.~E. Stamp.
\newblock Phase dynamics of solid-state qubits: Magnets and superconductors.
\newblock {\em Quantum Computers and Computing}, 4(1):20--62, 2003.

\bibitem{Belzig99}
W.~Belzig, F.K. Wilhelm, G.~Sch\"on, C.~Bruder, and A.D. Zaikin.
\newblock Quasiclassical {Green's} function approach to mesoscopic
  superconductivity.
\newblock {\em Superlattices and Microstructures}, 25:1251, 1999.

\bibitem{Rammer86}
J.~Rammer and H.~Smith.
\newblock Quantum field-theoretic methods in transport theory of metals.
\newblock {\em Review of Modern Physics}, 58(2):323, 1986.

\bibitem{Anthore03}
A.~Anthore, H.~Pothier, and D.~Esteve.
\newblock Density of states in a superconductor carrying a supercurrent.
\newblock {\em Phys. Rev. Lett.}, 90:127001, 2003.

\bibitem{Anderson58}
P.W. Anderson.
\newblock Absence of diffusion in certain random lattices.
\newblock {\em Phys. Rev.}, 109:1492, 1958.

\bibitem{Anthore03b}
A.~Anthore.
\newblock {\em Decoherence mechanisms in mesoscopic conductors}.
\newblock PhD thesis, Universite Paris 6, 2003.

\bibitem{Mooij09}
J.~E. Mooij.
\newblock private communication.

\bibitem{DunninghamRauBurnett2005}
J.~Dunningham, A.~Rau, and K.~Burnett.
\newblock From pedigree cats to fluffy-bunnies.
\newblock {\em Science}, 307(5711):872--875, Feb 2005.

\end{thebibliography}
\end{document}